\documentclass[prb,reprint,letterpaper,noeprint,longbibliography,nodoi]{revtex4-1} 

\usepackage{siunitx}
\usepackage{graphicx}
\usepackage{amsmath}
\usepackage[colorlinks, allcolors=blue]{hyperref}

\usepackage{placeins}

\begin{document}
\title{A Simple game simulating quantum measurements of qubits}
\author{Theodore A. Corcovilos}
\email{corcovilost@duq.edu}
\affiliation{Duquesne University, Dept.~of Physics, Pittsburgh, PA, 15282}
\affiliation{Pittsburgh Quantum Institute, Pittsburgh, PA 15260}
\date{\today}

\begin{abstract}\label{sec:abstract}
Games are useful tools for introducing new concepts to students.
This paper describes a competitive two-player game for sophomore students in a modern physics survey course or junior/senior students in an introductory quantum mechanics course to build intuition and quantitative understanding of the probabilistic nature of quantum measurements in two-level systems such as qubits or the Stern-Gerlach experiment.
The goal of the game is to guess a quantum state secretly chosen from a given set in the fewest number of measurements.
It uses twenty-sided dice or other classical random number generators to simulate quantum measurements.
The Bloch vector formalism is introduced to give a geometric description of the quantum states and measurement outcomes.
Several ready-to-use sets of quantum states are given, so readers can jump right in and try the game themselves without any prior knowledge of quantum mechanics.
More advanced students can also use the game in suggested follow-up exercises to deepen students' understanding of quantum measurements and their statistical description.
	
\end{abstract}

\maketitle
One of the challenges for students learning quantum mechanics is building intuition about the probabilistic nature of quantum measurements, particularly distinguishing between the result of a single measurement and the expectation value of that measurement.\cite{Zhu2012,Singh2015}
In classical physics, when we run an experiment multiple times with the same starting conditions, we expect to measure the same result each time, to within the uncertainty of our measuring devices.
However, when we repeat quantum mechanical experiments on identical systems many times, we generally obtain a statistical distribution of several possible measurement outcomes.
The ``quantum'' part of quantum mechanics comes from the way that certain physical quantities, such as angular momentum, can only take discrete values when we measure them.
The prototype quantum system we're considering here is a generic two-level system, a qubit, such as the spin of a spin-1/2 particle in a Stern-Gerlach--type experiment.\cite{Zhu2011}
Two-level quantum systems are key to many applications such as quantum computing,\cite{Strauch2016} nuclear magnetic resonance,\cite{Havel2002} quantum cryptography, and atomic clocks.
Two-level systems are also becoming more popular as the system of choice for introducing students to quantum mechanics.\cite{Townsend2012,McIntyre2012}
This ``spins-first'' approach aims to teach quantum axioms in  mathematically simpler finite Hilbert spaces before bringing in the complications of solving the wave equation.\cite{Kohnle2014,Sadaghiani2016}

This article describes: a simple two-level quantum system,  
how to simulate measurements on this system using a classical random number generator (RNG) such as dice rolls,
and a two-player competitive game using this simulation.
Although many excellent teaching-focused computer simulations of quantum mechanical systems,\cite{McKagan2008} and specifically two-level systems,\cite{Schroeder1993,Duer2014,Kohnle2015,Rycerz2015} exist, the exercise here takes a different approach.
Educational games are an effective way of engaging students.\cite{Goff2006}
We use a competitive game with dice to simulate single quantum measurements and demonstrate their probabilistic nature.
Dice are often used to demonstrate random processes in statistical mechanics\cite{Phillips2016,Timberlake2010} and quantum mechanics.\cite{Neto1984,Fleming2001}
The use of dice gives the students active participation in generating quantum data, rather than relying on unseen computer code to perform this function.
Also, using the dice slows down the process, giving students time for reflection, and emphasizes the importance of an individual quantum measurement.

The structure of the article follows.
After some brief comments about the learning objectives of this article (Sec.~\ref{sec:intro}), we begin with a brief introduction to two-level quantum systems using the Bloch vector formalism and describe the probabilistic nature of measurements on such a system (Sec.~\ref{sec:sys}).
Simulating quantum measurements on two-level systems using dice or other random number generators is described in Sec.~\ref{sec:sim}.
These concepts form the basis of a two-player competitive game, described in Sec.~\ref{sec:game}.
Readers who wish to play the game may skip directly to this section, which is self-contained.
Section \ref{sec:discussion} discusses one way to incorporate the game into coursework, informal observations of students, and some suggestions for follow-on exercises.
Appendix \ref{ap:spinors} contains a brief comparison of the Bloch vector and Dirac spinor formalisms, and Appendix \ref{ap:tables} contains additional ready-to-use game tables.

\section{Learning Objectives}\label{sec:intro}

The targets for these activities are junior- or senior-level introductory quantum mechanics students, particularly those in ``spins-first'' courses.\cite{Townsend2012,McIntyre2012,Kohnle2014,Sadaghiani2016}
The game also may be played before students learn the theory of quantum states and measurements as a qualitative introduction to these topics, for example as a teaser activity for a sophomore-level modern physics survey course\cite{Duer2014,Dur2016} or an introductory quantum computing course for non-physicists.\cite{Mermin2003}
Students who understand quantum measurements will also benefit because the core of the game is a foundational and subtle question: How much information can we gain from a single quantum measurement?

The learning objectives of the simulation and game described here are
\begin{itemize}
	\item To demonstrate the probabilistic nature of individual quantum measurements,
	\item To distinguish between the theoretical expectation values of measurements and the outcomes of individual measurements,
	\item To demonstrate how the statistical means of measurements approach the expectation values in the limit of many measurements,
	\item To distinguish between eigenstates of a measurement and non-eigenstates,
	\item To demonstrate that sets of incompatible measurements are needed to determine arbitrary quantum states,
	\item To build a qualitative feel for the amount of information contained in a single quantum measurement,
	\item For more advanced students, to provide a model system for practicing statistical estimation techniques,
	\item And lastly, to have fun learning quantum mechanics.
\end{itemize}

The key theme is understanding projective quantum measurements.
The students generate the measurement data themselves using classical random number generators.
Our random number generators of choice are twenty-sided dice.
These may be purchased at any gaming hobby shop or from internet vendors.
We use twenty-sided dice because they have the greatest number of sides among commonly available dice, giving us the highest precision for our random number generator.
Having only twenty possible measurement values causes a small amount of bias in the quantum measurement outcomes, but not enough to harm the pedagogical goals of the game.  
(If more precision is desired, one can generate outcomes to 1\% precision by using a pair of ten-sided dice with contrasting colors.
The sides of these are typically labeled 0--9. One die provides the ones digit of the percentage and the other die provides the tens digit, with a result of ``00'' representing 100\%.\cite{ADD1989}
For higher precision, one can use software to generate random numbers between 0 and 1 to any desired precision.)
The practical limitation of using dice for simulating quantum measurements is that it is inconvenient for simulating large numbers of measurements, although up to about twenty measurements can be done simultaneously without too much bother by rolling a large number of dice and sorting them into measurement outcomes.
For our game, however, we need only one measurement at a time.

We use the Bloch vector formalism of two-level systems as our mathematical framework,\cite{Bloch1946} but this is not essential if the instructor prefers the more common Dirac spinor notation.
A brief comparison of Bloch vectors and Dirac spinor notation for spin-1/2 states is given in Appendix \ref{ap:spinors} for interested readers.
Although often missing from introductory treatments of quantum mechanics, the Bloch picture is commonly used by practitioners of quantum computing,\cite{Chuang2001,Brun2002,Havel2002} and atomic clocks, among other things, because of its geometric interpretation and because the formalism is readily extended to describe open systems with dissipation, decoherence, and dephasing.\cite{Frimmer2014}
The Bloch vector for the spin angular momentum of a spin-1/2 particle responds to magnetic fields just like a classical magnetic moment does,
therefore students can use some of their classical intuition of rotations and torques to understand how spins respond to external forces and evolve in time.\cite{Supplee2000,Rojo2010,Wegrowe2012,Frimmer2014}
This is because of the formal definition of the Bloch vector in terms of expectation values (as we will show below in Eq.~\eqref{eq:sdef}) and Ehrenfest's Theorem, which states that these expectation values will obey classical physics when an appropriate classical analog exists.
For example, if we apply a constant magnetic field, the Bloch vector will precess about the magnetic field direction.
This makes the Bloch vector a useful tool for describing effects like nuclear magnetic resonance, for which it was first used.\cite{Bloch1946}
Similarly, quantum computing steps involving a single qubit, called single qubit gates, can be described as rotations of the qubit's Bloch vector.\cite{Chuang2001,Duer2014}
The Bloch vector picture of two-level systems also maps directly onto the Poincar\'e sphere picture of optical polarization,\cite{Dur2016,Jones2016} further strengthening the mathematical ties between classical and quantum physics.
The Bloch picture can also be interpreted using an analogy to coupled classical harmonic oscillators.\cite{Frimmer2014}
Lastly, Bloch vectors are useful when describing large ensembles of identical systems, such as macroscopic numbers of hydrogen nuclei in nuclear magnetic resonance measurements of solid state and chemical systems.\cite{Kittel2004,Engelhardt2015}

We're using the Bloch vector picture in this article because of the simple geometric interpretation of the states and the measurements in the hope that
because Bloch vectors are defined in a three-dimensional real-valued vector space, they provide an accessible geometric picture of spins states, ideal for students who are less experienced with complex-number algebra and linear algebra.
While the Dirac spinors are equivalent to Bloch vectors through the homomorphism from the $\mathrm{SU}(2)$ group of the Dirac spinors to the $\mathrm{SO}(3)$ group of the Bloch vectors,\cite{Tung1985} the geometric connection of the Dirac spinors to the measurement directions is less apparent than for the Bloch vectors: it is difficult to identify the expectation values of the spin measurements from the components of the Dirac spinors without doing a calculation, whereas for Bloch vectors a simple geometric projection suffices, as we will show in the next section.
Importantly, all of the concepts in this paper can be translated directly into the Dirac spinor formalism if one desires without changing the game described in Sec.~\ref{sec:game}.

\section{Two-state quantum systems in the Bloch vector picture}\label{sec:sys}
The mathematically simplest type of quantum system is one with only two possible basis states.
The system can be in either of these basis states or any linear superposition of them.
The canonical example of a two-level system is the Stern-Gerlach experiment with spin-$1/2$ atoms, which first demonstrated that the components of the angular momentum vector take on discrete values.\cite{Friedrich2003,Zhu2011}
Many other systems can be described or approximated as a two-state system: the spin angular momentum of an electron, the energy levels of atoms in an atomic clock, magnetic flux in a superconducting loop, electric charge on a quantum dot, or the states of electrons in the conduction and valence bands of a semiconductor.\cite{Dur2016}
One important application of two-level systems is to store information in quantum computers.
By assigning one state, for example spin in the $+z$ direction, to be 1 and the opposite, ``orthogonal,'' state, say spin in the $-z$ direction, to be 0, we have what is known as a quantum bit, or ``qubit.''
The process we're describing in this article and the basis of our game is measuring qubits, a vital step in any quantum computing algorithm.\cite{Chuang2001,Mermin2003}

\begin{figure}[h]
	\includegraphics[page=1]{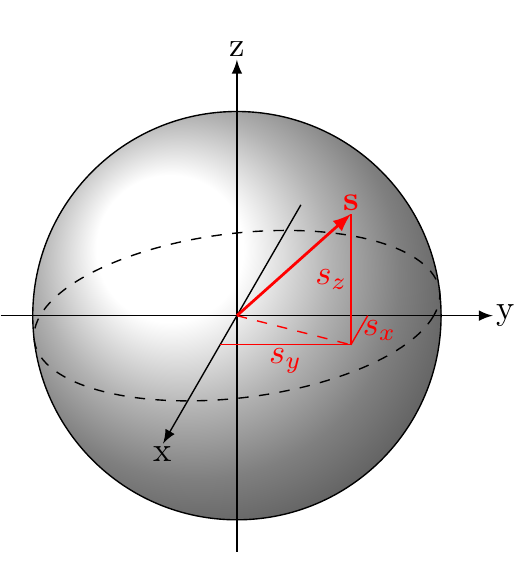}
	\caption{%
		(Color online)
		The Bloch vector $\mathbf{s}$ on the unit Bloch sphere.
		The Cartesian components of $\mathbf{s}$ are equal to the expectation values of measurements along the corresponding axis direction.
		The components of the vector obey $s_x^2+s_y^2+s_z^2 = 1$
		for pure states.%
	}\label{fig:bloch}
\end{figure}

We can describe the average direction of a spin-$1/2$ atom's or particle's spin angular momentum by a (real-valued) unit vector in three-dimensional space, called the Bloch vector, which we will represent by the vector $\mathbf{s}=(s_x,s_y,s_z)$ (Fig.~\ref{fig:bloch}).
All of the possible values of $\mathbf{s}$ for a single atom describe a unit-radius sphere, called the Bloch sphere.
We define $\mathbf{s}$ more rigorously below in Eq.~(\ref{eq:sdef}) of the next section and Eq.~\eqref{eq:diracS} of Appendix \ref{ap:spinors}, but interpreting it as the spin direction averaged over many measurements of identical systems is a good start.
Students should be cautioned however that Bloch vectors do not imply that the states of spin-1/2 systems are fixed in space.
Indeed, a primary goal of our game is to demonstrate that the
Bloch vectors are only obtained in the limit of a large number of measurements of identically prepared systems.

\begin{figure}[h]
	\includegraphics[page=2]{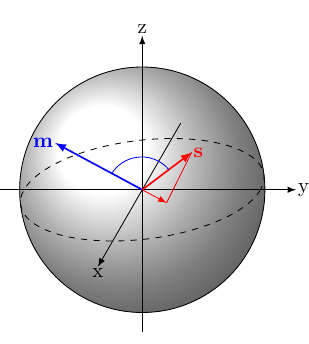}
	\caption{%
		(Color online)
		The measurement direction $\mathbf{m}$.
		The measurement probabilities are related to the angle (blue arc) between the vectors $\mathbf{s}$ and $\mathbf{m}$, or equivalently, the length of the projection of $\mathbf{s}$ onto $\mathbf{m}$ (small red arrow), through Eq.~(\ref{eq:probs}).%
	}\label{fig:meas}
\end{figure}

If we want to measure the spin vector, quantum mechanics says that we can only do a measurement along one direction at a time.
Let's label the measurement direction by a unit vector $\mathbf{m}$ (Fig.~\ref{fig:meas}).
The direction of $\mathbf{m}$ could be a physical direction (e.g.~the direction of the average magnetic field gradient in a Stern-Gerlach apparatus\cite{Zhu2011}) or a direction in an abstract mathematical space.
When we perform the measurement described by $\mathbf{m}$ on the state $\mathbf{s}$ the result is one of two possible values: $+1$ meaning that the spin after the measurement is along the direction of $\mathbf{m}$ or $-1$ meaning that the spin after the measurement is opposite the direction of $\mathbf{m}$.
(Following the suggestion of Ref.~\onlinecite{Mermin2003}, we are using units such that $\hbar/2=1$ to simplify the numerical values of the measurements -- our spin measurement operators are the Pauli operators and have eigenvalues $\pm1$.)
In the case where $\mathbf{s}$ is exactly along $\mathbf{m}$ or $-\mathbf{m}$, the result of the measurement will always be $+1$ or $-1$, respectively.  These values of $\mathbf{s}$ are called the \emph{eigenstates} of the measurement $\mathbf{m}$ with corresponding \emph{eigenvalues} of $\pm1$.

What happens if $\mathbf{s}$ is initially in some direction other than $\pm\mathbf{m}$?
Then we have a \emph{random chance} of measuring either $+1$ or $-1$.
The probability $P_m(\pm1)$ of a single measurement along $\mathbf{m}$ resulting in $+1$ or $-1$ depends on the angle between the $\mathbf{s}$ and $\mathbf{m}$ vectors:
\begin{equation}
\begin{aligned}
P_m(+1) &= \frac12 \left(1 + \mathbf{m}\cdot\mathbf{s} \right), \\
P_m(-1) &= \frac12 \left(1 - \mathbf{m}\cdot\mathbf{s} \right),
\end{aligned}\label{eq:probs}
\end{equation}
where $\cdot$ represents the vector dot product.
(The derivation of this result is given in Appendix~\ref{ap:spinors} leading to Eq.~\eqref{eq:probsD}.)
Note that these probabilities add up to one,
and in the case of $\mathbf{s}$ being an eigenstate of $\mathbf{m}$ we get probabilities of 0 or 1.
Only the eigenstates of the measurement $\mathbf{m}$ can be measured with zero uncertainty.

The type of measurement we're describing is called a projective measurement because the measurement projects the spin into a final state after the measurement of either $+\mathbf{m}$ or $-\mathbf{m}$, depending on the outcome of the measurement.
This projection process, in which the measurement changes the state of the system being measured, prevents subsequent measurements from revealing additional information about the initial state $\mathbf{s}$.
To gain more information, we must reset the experiment by returning the system to its initial condition (or generate a new system in an identical initial state) before performing another measurement.
Alternatively, we could prepare a large ensemble of identical systems and measure the ensemble averages of the measurement -- this is the usual practice in nuclear magnetic resonance and atomic clock experiments.

If we repeat the measurement many times, each time reseting the system to state $\mathbf{s}$ before the measurement, the expected mean value of the measurements is
\begin{equation}
\label{eq:ev}
\bar{P}_m = P_m(+1) - P_m(-1) = \mathbf{m}\cdot\mathbf{s}.
\end{equation}
In particular, if we calculate the expectation values of the measurements along the directions $\mathbf{m}=x$, $y$, and $z$, we get the components of the vector $\mathbf{s}$ itself:
\begin{equation}
\mathbf{s} = \left(\bar{P}_x,\bar{P}_y,\bar{P}_z  \right).\label{eq:sdef}
\end{equation}
The above equation is the formal definition of the Bloch vector $\mathbf{s}$.

\section{A simple simulation of quantum measurements}\label{sec:sim}
Quantum measurements of a two-level system such as the spin of a spin-1/2 atom or the state of a qubit can be simulated with a classical random number generator 
using the probabilities in Eq.~\eqref{eq:probs}.
Our favorite random number generator is a twenty-sided die,
but a computer's/calculator's random number generator or smartphone application can be used as well.

To convert a die roll into the result of a measurement in the $\mathbf{m}$ direction (which is not necessarily along a coordinate axis):
\begin{enumerate}
	\item Roll a 20-sided die and multiply the result by $5\%$, or generate a random number between 0 and 1 and multiply it by $100\%$.
	\item If this percentage is \emph{less than or equal to} $P_m(+1)$, as calculated in Eq.~\eqref{eq:probs}, then the result of the measurement is $+1$. If this percentage is \emph{greater than} $P_m(+1)$ then the result of the measurement is $-1$.
\end{enumerate}
The possible measurement outcomes for a state with $P_m(+1) = 70\%$ are depicted in Table~\ref{tab:bar}.

\begin{table}
	\caption{Example of measurement outcomes in the case $P_m(+1) = 70\%$}\label{tab:bar}
	\centering%
	\begin{tabular}{cc|c}
		20-Sided die roll & \phantom{M}RNG\phantom{M} & Result \\ 
		\hline
		0--14 & $\le$ 70\% & $+1$ \\
		15--20 & $>$70\% & $-1$
	\end{tabular}
\end{table}

In quantum experiments a single measurement by itself is not sufficient to determine the state $\mathbf{s}$.
We need to perform many measurements and compile statistics.
To estimate the initial state of the system we can create many copies of the system and perform a measurement on each copy
(e.g.~roll the die many times or roll many dice simulataneously).
If we repeat the measurement along direction $\mathbf{m}$ $N$ times and get $N_+$ results of $+1$ and $N_-$ results of $-1$, then the mean value of the measurement along this direction is
\begin{equation}
\left<{P}_m\right> = \frac{N_+-N_-}{N},\label{eq:mean}
\end{equation}
and the statistical uncertainty of the measurements for large $N$ is the sample deviation:\cite{Wallis2013}
\begin{equation}
\sigma_m = \sqrt{\frac{1-\left<{P}_m\right>^2}{N-1}}.\label{eq:sig}
\end{equation}
For large enough $N$, the measurement mean $\left<P_m\right>$ approaches the expectation value $\bar{P}_m$ given in Eq.~\eqref{eq:ev}.%
\footnote{%
Statistically speaking, the simulation represents three independent Bernoulli processes with probabilities $P_x(+1)$, $P_y(+1)$, and $P_z(+1)$, where the values of the probabilities are related to each other by the normalization constraint that the length of the Bloch vector equals one.
This constraint complicates the problem of estimating the state $\mathbf{s}$ from measurement data -- na{{\"i}}ve application of Eqs.~(\ref{eq:mean}) and (\ref{eq:sig}) can yield nonphysical estimates of the state that violate the normalization constraint $s_x^2+s_y^2+s_z^2 = 1$.
}
A separate set of measurements is needed for each of three linearly independent directions $\mathbf{m}$ (e.g.~$x$, $y$, and $z$) to unambigously determine the state $\mathbf{s}$.

When presenting these concepts to students, I typically introduce them by giving the students various combinations of $\mathbf{s}$ and $\mathbf{m}$ and asking them to simulate $N=10$ or so measurements following the instructions at the top of this section,
enough to show the convergence of the means towards the expectation values.
After the students have had practice simulating the measurements, I'll ask them how many measurements are needed to get a ``good'' estimate of the expectation value.
They usually respond with some large number like one thousand or one million.  A student familiar with statistics may point out that the measurement uncertainty goes like $1/\sqrt{N}$, as in Eq.~\eqref{eq:sig}.
Because $N$ needs to be large to precisely determine the states, why would we use a slow process such as dice rolls to simulate measurements?

Indeed, several computer simulations of quantum measurements exist,\cite{Schroeder1993,Duer2014,Kohnle2015,Rycerz2015} allowing students to simulate thousands of measurements nearly instantly to get a precise measurement mean.
While a computer is useful for generating large sample sizes, automation obscures the granular way in which the data develop.
Slowing down the measurement process by using a mechanical method, such as dice rolling, lets the students experience and appreciate the randomness inherent in quantum processes and how the mean values converge over many measurements to the expectation value.

To motivate discussion about why understanding individual measurements is important, it is useful to ask the students,
``What would you do if each measurement was really expensive?''
and,
``How would you decide when you have enough data to known the initial state?''
These questions inspired the game described in the next section.

More concepts can be added for advanced students once the mechanics of the simulation are understood.
For example, incoherent mixtures of states can be described by Bloch vectors $\mathbf{s}_{\text{mix}}$ with length less than one.
The incoherent mixture of states $\mathbf{s}_i$, each contributing fraction $p_i$ of the population, has a Bloch vector
\begin{equation}\label{eq:mixed_s}
\mathbf{s}_{\text{mix}} = \sum\limits_{i} p_i \mathbf{s}_i.
\end{equation}
Measurement results of these mixtures are given by Eq.~\eqref{eq:probs} and the expectation values given by Eq.~\eqref{eq:ev} with $\mathbf{s}$ replaced by $\mathbf{s}_{\text{mix}}$.
In particular, the Bloch vector for a completely incoherent mixture is the zero vector -- a measurement in any direction has an equal chance of returning $+1$ or $-1$.
Similarly, imperfect measurements can be simulated by letting the measurement vectors $\mathbf{m}$ have length less than one, representing a chance of getting an incorrect measurement result, even for eigenstates.

\section{The Quantum State Guessing Game}\label{sec:game}
The simulation described above can be used as the basis for a competitive two-player game encompassing the concepts of quantum measurement.
The theme of the game is to determine the state $\mathbf{s}$ of a two-level quantum system from a given set of possible states using the fewest measurements.
The two players are called the Scientist and the Experiment.
The goal of the Scientist is to determine a secret state using the fewest number of measurements.
The Experiment chooses the secret state, conducts the measurements, and reports the results.
The game requires a random number generator (such as a twenty-sided die or computer/calculator/smartphone) and a table of probabilities of the measurement outcomes for each possible state, which I'll call the ``game table.''

\begin{table}[h!]
	\caption{%
		Example game table: Eigenstates (Color online)%
	}\label{tab:game1}
	{\centering
		\includegraphics[page=1,scale=0.74]{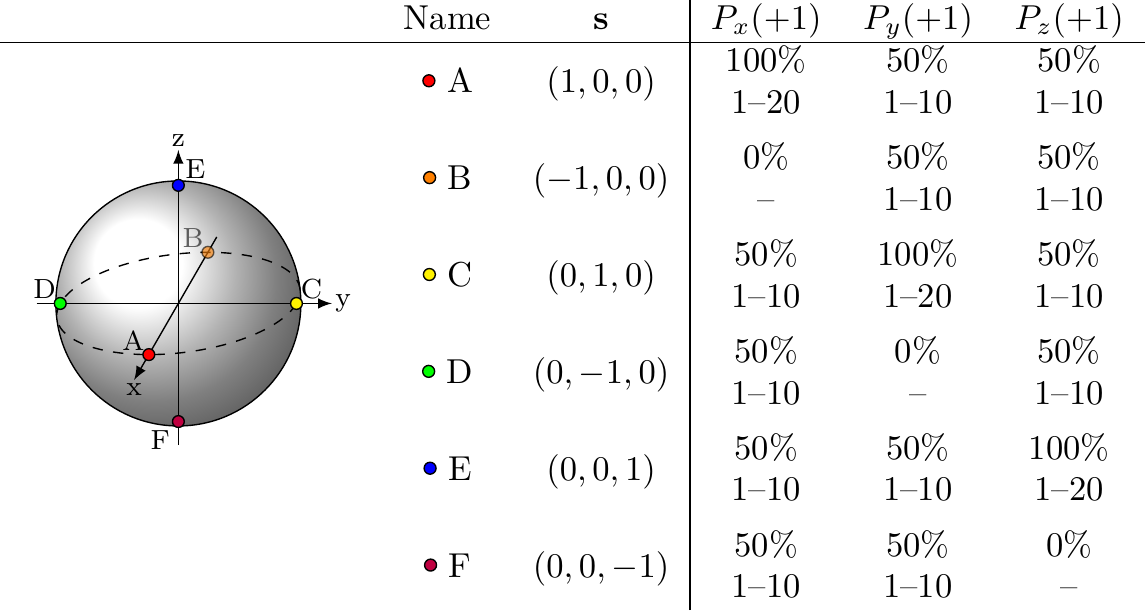}	
	}
\end{table}

The game table 
(such as Table~\ref{tab:game1} or those in Appendix \ref{ap:tables})
shows a predetermined set of possible state Bloch vectors $\mathbf{s}$ and the probabilities of measuring $+1$ from measuring the spin in the $x$, $y$, or $z$ directions.
The range of twenty-sided die roll outcomes corresponding to the $+1$ measurement result (see top of Sec.~\ref{sec:sim} for how these are calculated) are also shown below the percentages to speed up the measurement simulation.
If the die roll is outside the range shown, the result of the measurement is $-1$.
Both players may consult the game table while playing the game.

A table for any set of states may be generated by the instructor or the students using Eq.~\eqref{eq:probs} to find the probability of measuring $+1$ for the measurement direction $\mathbf{m}$ equal to the $x$, $y$, and $z$ unit vectors.
To convert percentages to dice values, divide the percentage by the number of sides on the dice (e.g.~20), and round to the nearest integer, which may be zero.
This value is the maximum dice roll that yields a $+1$ measurement.
Any die roll greater than this value yields a $-1$ measurement.

Play proceeds as follows:
\begin{enumerate}
	\item To begin, the Experiment secretly picks a state $\mathbf{s}$ from the list of states in the game table.
	\item The Scientist asks for a measurement of spin in either the $x$, $y$, or $z$ direction.\label{it:meas}
	\item The Experiment secretly rolls the twenty-sided die 
	and consults the table entry corresponding to the row of the secret state and the column of the requested measurement direction.
	\begin{enumerate}
		\item If the roll result is within the range shown in the game table, the Experiment reports a result of $+1$ to the Scientist.
		\item If the roll result is outside the range shown in the game table, the Experiment reports $-1$.
	\end{enumerate}
	\item At this point, the Scientist may try to guess the secret state or may request a new measurement in any direction (going back to step \ref{it:meas}).
	\begin{enumerate}
		\item If the Scientist guesses incorrectly, they receive a penalty of $+5$ points.
		The Scientist may either guess again or ask for a new measurement.
		\item If the Scientist guesses correctly, the round ends and the Scientist's score is the number of measurements performed plus any penalties for bad guesses.
	\end{enumerate}
\end{enumerate}
Like golf, a lower score is better.  
An example score card is shown in Fig.~\ref{fig:score}.
After the Scientist correctly guesses the state, the players swap roles and play again.
After several rounds (three per player is a good number), the player with the lowest total score wins!

\begin{figure}[h]
{
\resizebox{3.4in}{!}{%
\centering

\begin{tabular}{c|ccc|l}
	Measurement & Direction  & Die roll & Result & Guess \\ 
	Number & (Scientist) & (Exp., Secret) & (Experiment) & (Scientist) \\ \hline
	1 & {X} & {7} & {+1} &                                   \\
	2 & {X} & {12}& {-1} &                                    \\
	3 & {Z} & {15}& {-1} &                                    \\
	4 & {Z} & {4} & {-1} &                                   \\
	5 & {Y} & {1} & {+1} &                                   \\
	6 & {Y} & {8} & {+1} & {C} \emph{Wrong! +5}\\
	7 & {Z} & {10}& {-1} & {F} \emph{Correct!}  \\
	\multicolumn{5}{c}{\normalsize\emph{Score:} $7+{5}=12$}
\end{tabular}
}
}
\caption{%
	An example scorecard for the system in Table~\ref{tab:game1} for the case where the the unknown state chosen by the Experiment player is the ``F'' state, $\mathbf{s} = (0,0,-1)$.
	Each turn the Scientist player requests a measurement direction ($\mathbf{m}=x$, $y$, or $z$).
	The Experiment rolls the die (center column, hidden from the Scientist) and consults Table~\ref{tab:game1} to give a result of $+1$ or $-1$.
	The Scientist may guess the state at any time, but an incorrect guess results in a $+5$ point penalty.
	In the game shown, the Scientist used 7 measurements and made one incorrect guess of the state, giving a total score of 12.
	A lower score is better.%
}\label{fig:score}
\end{figure}

\section{Discussion}\label{sec:discussion}
The states chosen for Table~\ref{tab:game1} are the eigenstates of the measurement directions.
The key property of an eigenstate of a measurement is that the corresponding measurement always yields the same outcome (always $+1$ or always $-1$).
The students soon discover that for this table they can deduce the secret state by a process of elimination.
For example, if any measurement of $y$ yields a result of $+1$, then the state $(0,-1, 0)$ is not possible.
Consulting the game table shows that the corresponding entry reads $0\%$.
After the students are comfortable with the concept of eigenstates, they then should try a game table without measurement eigenstates, such as Tables~\ref{tab:game2}--\ref{tab:game4} in Appendix \ref{ap:tables}.
Without eigenstates, strict elimination is not viable, so the students must evaluate the results probabilistically and more carefully consider when to guess the state.

Informal observations of the students%
\footnote{%
The students were  six sophomore students in a Modern Physics survey course and eleven junior/senior students in an Introductory Quantum Mechanics course taught from the spins-first perspective using Ref.~7 as the primary textbook.
}
while playing the game show the students were actively engaged.
The students were talkative and bought into the competitive nature of the game.
It was entertaining for the instructor to watch as the students developed strategies for playing the game.
Early on they used a simple strategy of eliminating states based on single measurement results, as if every state were a measurement eigenstate.
Then they became more skeptical of any single result because of the chance of ``bad'' (unlikely) measurements.
After a couple of rounds the students developed more sophisticated strategies, as evidenced by more accurate and more aggressive guessing, resulting in better scores.
However, when asked to explain their strategies, most had difficultly articulating how they were choosing measurements or deciding when to guess.
It was encouraging to see the students develop some intuitive understanding about the quantum measurement process.

To reinforce the students' conceptual thinking gained while playing the game with more rigorous quantitative understanding, we gave the advanced students a follow-on project
to write a computer program to play the game.
Writing the Experiment player is relatively simple and gives the students a chance to practice using random numbers to generate quantum measurement results.
Writing a good algorithm for the Scientist role, however, is quite challenging.
Advanced undergraduates may know some tools of statistical estimation that they can use to interpret the measurement data.
	Some methods used by my students were the method of moments using Eqs.~(\ref{eq:mean}) and (\ref{eq:sig}), the maximum likelihood method, and Bayesian estimation.
(Derivations and technical comparisons of these statistical methods are beyond the scope of this article.)
Deciding which state is most likely is straightforward using statistical methods,
but deciding which measurement to make next or when to guess is not obvious and can lead to more in-depth discussion about the nature of quantum information.

\section{Conclusion}\label{sec:conclusion}
Here we presented the rules governing projective quantum measurements of two-level systems and used these rules as the basis for a simple, but deep, two-player competitive game to build student intuition and statistical reasoning about quantum measurements.
The game requires students to consider how much information is necessary to identify a quantum state.
Specific topics addressed include measurement probabilities, the distinction between individual measurements and expectation values, and the special nature of measurement eigenstates.

\appendix
\section{Dirac spinor notation}\label{ap:spinors}
We have used the Bloch vector formalism in the main text because of its geometric connection (e.g.~Fig.~\ref{fig:bloch}) and because beginning students are more comfortable with real-valued Cartesian vectors.
The measurements simulated in the text may also be described using Dirac spinor notation.
The key results appear below without proof for readers already familiar with spinor notation; these are found in most standard textbooks, such as Refs.~\onlinecite{Griffiths2004,McIntyre2012,Townsend2012}.
Spinors are of course the more ``professional'' description of quantum mechanics because they can be directly extended to systems with more than two basis states.

If the basis states are chosen to be the $z$-direction eigenstates, denoted
\begin{equation}\label{eq:z-eig}
\vec{\phi}_{+z} = \binom{1}{0}
\text{ and }
\vec{\phi}_{-z} = \binom{0}{1},
\end{equation}
then an arbitrary pure state may be written
\[
\vec{\psi} = a \vec{\phi}_{+z} + b\vec{\phi}_{-z} = \binom{a}{b},
\]
where $a$ and $b$ are complex numbers and obey the normalization constraint
$ |a|^2 + |b|^2 = 1$.
The spin direction measurement operators are represented by the Pauli matrices
\begin{equation}\label{eq:pauli}
\sigma_z = \begin{pmatrix} 1 & 0 \\ 0 & -1 \end{pmatrix},
\sigma_x = \begin{pmatrix} 0 & 1\\ 1 & 0 \end{pmatrix},
\text{ and }
\sigma_y = \begin{pmatrix} 0 & -i\\ i & 0 \end{pmatrix},
\end{equation}
with $i=\sqrt{-1}.$
These each have eigenvalues of $\pm1$. The eigenvectors of $\sigma_x$ and $\sigma_y$ are, respectively,
\[
\vec{\phi}_{\pm x} = \frac{1}{\sqrt{2}}\,\binom{1}{\pm1},
\vec{\phi}_{\pm y}  = \frac{1}{\sqrt{2}}\,\binom{1}{\pm i},
\]
with the eigenvectors of $\sigma_z$ given above in Eq.~\eqref{eq:z-eig}.

The probability of a measurement in the $j= x,y,z$ direction returning a value of $\pm1$ is given by the Born rule using the eigenvectors of the $\sigma_j$ operators,
\[
P_j(\pm1) = \left|\vec{\psi}^{\,\dagger} \vec{\phi}_{\pm j} \right|^2,
\]
where $\dagger$ indicates the conjugate transponse: $\vec{\psi}^{\,\dagger} = (a^*, b^*)$, a row vector, with $*$ being the complex conjugate.
In particular,
\begin{equation}
\begin{gathered}
P_x(\pm1) = \frac12 \left(1\pm 2\, \mathrm{Re}[ab^*] \right), \\
P_y(\pm1) = \frac12 \left(1\pm 2\, \mathrm{Im}[ab^*] \right), \\
P_z(+1) = |a|^2,\quad P_z(-1) = |b|^2,
\end{gathered}
\end{equation}
where Re and Im give the real part and imaginary part, respectively.

Using these measurement probabilities we can construct the expectation values:
\[
\bar{P}_j = P_j(+1) - P_j(-1) = \vec{\psi}^{\,\dagger}\sigma_j\vec{\psi},
\]
\begin{equation}\label{eq:spinorprobs}
\begin{aligned}
\bar{P}_x &= 2\,\mathrm{Re}[ab^*], \\
\bar{P}_y &= 2\,\mathrm{Im}[ab^*], \\
\bar{P}_z &= |a|^2 - |b|^2.
\end{aligned}
\end{equation}
The primary motivation for using the Bloch vector formalism is the contrast between Eq.~\eqref{eq:spinorprobs} and Eq.~\eqref{eq:ev} in the main text:
\begin{equation}
\tag{\ref{eq:ev}}
\bar{P}_m = P_m(+1)-P_m(-1)=\mathbf{m}\cdot\mathbf{s},
\end{equation}
where $\mathbf{m}$ is an arbitrary measurement direction.
The Bloch version is more compact and arguably easier to interpret.

Using the definition of the Bloch vector $\mathbf{s}$ in terms of the expectation values, Eq.~\eqref{eq:sdef}, we get the correspondence between the spinor $\vec{\psi}$ and $\mathbf{s}$:
\begin{equation}\label{eq:diracS}
\mathbf{s} = \left(2\, \mathrm{Re}[ab^*], 2\,\mathrm{Im}[ab^*], |a|^2-|b|^2 \right).
\end{equation}
Conversely, a unit-length Bloch vector expressed in spherical coordinates
\[
\varphi = \arctan(s_y/s_x),
\text{ and }
\theta = \arctan\left(\sqrt{s_x^2+s_y^2}\,/s_z\right),
\]
has a spinor representation of 
\[
\vec{\psi} = \binom{\cos(\theta/2)}{\sin(\theta/2)e^{i\varphi}}.
\]

A spin measurement along arbitrary direction unit-vector $\mathbf{m}$, not necessarily along the coordinate axes, corresponds to a measurement operator
\begin{equation}
\label{eq:sigmam}
\sigma_m = m_x\sigma_x + m_y\sigma_y + m_z \sigma_z = \begin{pmatrix}
m_z & m_x-im_y \\ m_x+im_y & -m_z
\end{pmatrix},
\end{equation}
which has eigenvalues $\pm 1$ and corresponding eigenvectors
\[
\vec{\phi}_{+m} = \binom{\cos(\alpha/2)}{\sin(\alpha/2)e^{i\beta}}
,\qquad 
\vec{\phi}_{-m} = \binom{\sin(\alpha/2)}{-\cos(\alpha/2)e^{i\beta}},
\]
where the angles are defined by
\[
\beta = \arctan(m_y/m_x),
\text{ and }
\alpha = \arctan\left(\sqrt{m_x^2+m_y^2}\,/m_z\right).
\]
The measurement probabilities for the direction $\mathbf{m}$ are
\begin{align*}
P_m(+1) &= 
\left|\vec{\psi}^{\,\dagger}\phi_{+m}\right|^2 \\
&= \cos^2(\theta/2)\cos^2(\alpha/2) + \sin^2(\theta/2)\sin^2(\alpha/2)  \\
&\qquad +2\cos(\theta/2)\cos(\alpha/2)\sin(\theta/2)\sin(\alpha/2)\cos(\varphi-\beta). 
\end{align*} 
Using trigonometric identities and the definitions of the angles, this simplifies to
\begin{equation}\label{eq:probsD}
\begin{aligned}
P_m(+1)&= \frac12 (1+s_x m_x+ s_y m_y + s_z m_z), \\
P_m(-1)&= \frac12 (1-s_x m_x- s_y m_y - s_z m_z),
\end{aligned}  
\end{equation}
yielding Eq.~\eqref{eq:probs} in the main text.

Incoherent mixtures of states are given by Bloch vectors with length $s_x^2+s_y^2+s_z^2<1$, as shown in Eq.~\eqref{eq:mixed_s}.
These mixtures cannot be expressed as single spinors, but lead to a density matrix of
\[
\rho = \frac12 \left(s_x\sigma_x + s_y\sigma_y + s_z\sigma_z + I\right) =
\frac12
\begin{pmatrix}
1 + s_z & s_x-is_y \\ s_x+is_y & 1-s_z
\end{pmatrix},
\]
where the Pauli matrices are given by Eq.~(\ref{eq:pauli}) and $I$ is the identity matrix.
The expectation value of a spin measurement $\sigma_m$, from Eq.~\eqref{eq:sigmam}, is given by
\[
\bar{P}_m = \mathrm{Trace}[\rho \sigma_m],
\]
which also simplifies to Eqs.~\eqref{eq:probs} and \eqref{eq:probsD}.

\section{Additional game tables}\label{ap:tables}

Some additional example game tables appear here.
The sets of states for each of these example games tables were chosen to be symmetrically placed on the Bloch sphere and oriented to give subjectively interesting ranges of measurement outcomes.
For these examples the states are the vertices of regular polyhedra, but any set of states with sufficient separation may be used.
A set of six or eight random states can stimulate discussion and experimentation of optimal measurement strategies, or as test cases for game-playing algorithms.
Games with more than eight possible states tend to converge slowly and become tedious for human players, but provide interesting test scenarios for computer players.

\begin{table}[h!]
	\caption{Game table: Tilted tetrahedron (Color online)}\label{tab:game2}
	{\centering
		\includegraphics[page=2,scale=0.74]{QG-tabs}	
	}
\end{table}

\begin{table}[h!]
	\caption{Game table: Tilted octohedron (Color online)}\label{tab:game3}
	{\centering
		\includegraphics[page=4,scale=0.74]{QG-tabs}	
	}
\end{table}

\begin{table}[h!]
	\caption{Game table: Cube (Color online)}\label{tab:game4}
	{\centering
		\includegraphics[page=5,scale=0.74]{QG-tabs}	
	}
\end{table}

\begin{acknowledgments}
	The author thanks M.~Huster, C.~Singh, J.~Kern, E.~Hazlett, N.~Lundblad, and D.~Schroeder for helpful discussions,
	and the students of the author's Physics 473 class (Fall 2016) and Y.~Wang's Physics 321 class (Spring 2017) for participating in the activities and providing feedback to improve the game.
	Travel support for discussions while writing this article was provided by the Bayer School for Natural and Environmental Sciences at Duquesne University through the ``Entering Mentoring'' program.
\end{acknowledgments}

\FloatBarrier


\begin{thebibliography}{10}
	\newcommand{\enquote}[1]{``#1''}
	
	\bibitem{Zhu2012}
	Guangtian Zhu and Chandralekha Singh, \enquote{Improving students'
	  understanding of quantum measurement. {I.} {Investigation} of difficulties,}
	  \href{http://dx.doi.org/10.1103/PhysRevSTPER.8.010117}{Physical Review
	  Special Topics - Physics Education Research} \textbf{8}~(1), 010117 (2012).
	\bibitem{Singh2015}
	Chandralekha Singh and Emily Marshman, \enquote{Review of student difficulties
	  in upper-level quantum mechanics,}
	  \href{http://dx.doi.org/10.1103/PhysRevSTPER.11.020117}{Physical Review
	  Special Topics - Physics Education Research} \textbf{11}~(2), 1--24 (2015),
	  and references therein.
	\bibitem{Zhu2011}
	Guangtian Zhu and Chandralekha Singh, \enquote{Improving students'
	  understanding of quantum mechanics via the {Stern--Gerlach} experiment,}
	  \href{http://dx.doi.org/10.1119/1.3546093}{American Journal of Physics}
	  \textbf{79}~(5), 499--507 (2011).
	\bibitem{Strauch2016}
	Frederick~W. Strauch, \enquote{Resource letter {QI}-1: {Quantum} information,}
	  \href{http://dx.doi.org/10.1119/1.4948608}{American Journal of Physics}
	  \textbf{84}~(7), 495--507 (2016), particularly, Section {II}.
	\bibitem{Havel2002}
	T.~F. Havel \emph{et~al.}, \enquote{Quantum information processing by nuclear
	  magnetic resonance spectroscopy,}
	  \href{http://dx.doi.org/10.1119/1.1446857}{American Journal of Physics}
	  \textbf{70}~(3), 345--362 (2002).
	\bibitem{Townsend2012}
	John~S. Townsend, \emph{A Modern Approach to Quantum Mechanics}, 2nd edition
	  (University Science Books, Mill Valley, 2012).
	\bibitem{McIntyre2012}
	David~H. McIntyre, Corrine~A. Manogue, and Janet Tate, \emph{Quantum Mechanics:
	  {A} Paradigms Approach} (Pearson Education, San Fransisco, 2012).
	\bibitem{Kohnle2014}
	Antje Kohnle \emph{et~al.}, \enquote{A new introductory quantum mechanics
	  curriculum,}
	  \href{http://stacks.iop.org/0143-0807/35/i=1/a=015001?key=crossref.b893bf95459700d33f0d15b175edcfc6}{European
	  Journal of Physics} \textbf{35}~(1), 015001 (2014).
	\bibitem{Sadaghiani2016}
	Homeyra~R. Sadaghiani, \enquote{Spin first vs. position first instructional
	  approaches to teaching introductory quantum mechanics,} in \emph{2016
	  Physics Education Research Conference Proceedings}, pp. 292--295 (American
	  Association of Physics Teachers, 2016).
	\bibitem{McKagan2008}
	S.~B. McKagan \emph{et~al.}, \enquote{Developing and researching {PhET}
	  simulations for teaching quantum mechanics,}
	  \href{http://dx.doi.org/10.1119/1.2885199}{American Journal of Physics}
	  \textbf{76}~(2008), 406--417 (2008).
	\bibitem{Schroeder1993}
	Daniel~V. Schroeder and Thomas~A. Moore, \enquote{A computer-simulated
	  {Stern--Gerlach} laboratory,}
	  \href{http://dx.doi.org/10.1119/1.17172}{American Journal of Physics}
	  \textbf{61}~(9), 798--805 (1993).
	\bibitem{Duer2014}
	Wolfgang D{\"{u}}r and Stefan Heusler, \enquote{Visualization of the invisible:
	  {T}he qubit as key to quantum physics,}
	  \href{http://dx.doi.org/10.1119/1.4897588}{Physics Teacher} \textbf{52}~(8),
	  489--492 (2014).
	\bibitem{Kohnle2015}
	Antje Kohnle, Charles Baily, Anna Campbell, and Natalia Korolkova,
	  \enquote{Enhancing student learning of two-level quantum systems with
	  interactive simulations,} \href{http://dx.doi.org/10.1119/1.4913786}{American
	  Journal of Physics} \textbf{83}~(6), 560--566 (2015).
	\bibitem{Rycerz2015}
	Katarzyna Rycerz, Joanna Patrzyk, Bart{{\l}}omiej Patrzyk, and Marian Bubak,
	  \enquote{Teaching quantum computing with the {QuIDE} simulator,}
	  \href{http://dx.doi.org/10.1016/j.procs.2015.05.374}{Procedia Computer
	  Science} \textbf{51}, 1724--1733 (2015).
	\bibitem{Goff2006}
	Allan Goff, \enquote{Quantum tic-tac-toe: {A} teaching metaphor for
	  superposition in quantum mechanics,}
	  \href{http://dx.doi.org/10.1119/1.2213635}{American Journal of Physics}
	  \textbf{74}~(11), 962--973 (2006).
	\bibitem{Phillips2016}
	Jeffrey~A. Phillips, \enquote{The macro and micro of it is that entropy is the
	  spread of energy,} \href{http://dx.doi.org/10.1119/1.4961175}{Physics
	  Teacher} \textbf{54}~(6), 344--347 (2016).
	\bibitem{Timberlake2010}
	Todd Timberlake, \enquote{The statistical interpretation of entropy: {An}
	  activity,} \href{http://dx.doi.org/10.1119/1.3502501}{Physics Teacher}
	  \textbf{48}~(8), 516--519 (2010).
	\bibitem{Neto1984}
	Benicio de~Barros~Neto, \enquote{Dice throwing as an analogy for teaching
	  quantum mechanics,} \href{http://dx.doi.org/10.1021/ed061p1044}{Journal of
	  Chemical Education} \textbf{61}~(12), 1044--1045 (1984).
	\bibitem{Fleming2001}
	Patrick~E. Fleming, \enquote{A quantum mechanical game of craps: {Teaching} the
	  superposition principle using a familiar classical analog to a quantum
	  mechanical system,} \href{http://dx.doi.org/10.1021/ed078p57}{Journal of
	  Chemical Education} \textbf{78}~(1), 57--60 (2001).
	\bibitem{Dur2016}
	Wolfgang D{\"{u}}r and Stefan Heusler, \enquote{The qubit as key to quantum
	  physics {Part II}: {Physical} realizations and applications,}
	  \href{http://dx.doi.org/10.1119/1.4942137}{Physics Teacher} \textbf{54}~(3),
	  156--159 (2016).
	\bibitem{Mermin2003}
	N.~David Mermin, \enquote{From {Cbits} to {Qbits}: {Teaching} computer
	  scientists quantum mechanics,}
	  \href{http://dx.doi.org/10.1119/1.1522741}{American Journal of Physics}
	  \textbf{71}~(1), 23--30 (2003).
	\bibitem{ADD1989}
	David Cook and Mike Breault, \emph{Advanced Dungeons \& Dragons: Player's
	  Handbook}, 2nd edition ({TSR}, Lake Geneva, 1989).
	\bibitem{Bloch1946}
	F.~Bloch, \enquote{Nuclear induction,}
	  \href{http://dx.doi.org/10.1103/PhysRev.70.460}{Physical Review}
	  \textbf{70}~(7-8), 460--474 (1946).
	\bibitem{Chuang2001}
	Isaac~L. Chuang and Michael~A. Nielsen, \emph{Quantum Computation and Quantum
	  Information} (Cambridge, Cambridge, 2001).
	\bibitem{Brun2002}
	Todd~A. Brun, \enquote{A simple model of quantum trajectories,}
	  \href{http://dx.doi.org/10.1119/1.1475328}{American Journal of Physics}
	  \textbf{70}~(7), 719--737 (2002).
	\bibitem{Frimmer2014}
	Martin Frimmer and Lukas Novotny, \enquote{The classical {Bloch} equations,}
	  \href{http://dx.doi.org/10.1119/1.4878621}{American Journal of Physics}
	  \textbf{82}~(10), 947--954 (2014).
	\bibitem{Supplee2000}
	James~M. Supplee, \enquote{Optical {Bloch} equations: {Informal} motivation
	  without the {Schr{\"{o}}dinger} equation,}
	  \href{http://dx.doi.org/10.1119/1.19392}{American Journal of Physics}
	  \textbf{68}~(2), 180--185 (2000).
	\bibitem{Rojo2010}
	Alberto~G. Rojo and Anthony~M. Bloch, \enquote{The rolling sphere, the quantum
	  spin, and a simple view of the {Landau}--{Zener} problem,}
	  \href{http://dx.doi.org/10.1119/1.3456565}{American Journal of Physics}
	  \textbf{78}~(10), 1014--1022 (2010).
	\bibitem{Wegrowe2012}
	J.-E. Wegrowe and M.-C. Ciornei, \enquote{Magnetization dynamics, gyromagnetic
	  relation, and inertial effects,}
	  \href{http://dx.doi.org/10.1119/1.4709188}{American Journal of Physics}
	  \textbf{80}~(7), 607--611 (2012).
	\bibitem{Jones2016}
	Joshua~A. Jones \emph{et~al.}, \enquote{The {Poincar{\'{e}}}-sphere approach to
	  polarization: {Formalism} and new labs with {Poincar{\'{e}}} beams,}
	  \href{http://dx.doi.org/10.1119/1.4960468}{American Journal of Physics}
	  \textbf{84}~(11), 822--835 (2016).
	\bibitem{Kittel2004}
	Charles Kittel, \emph{Introduction to Solid State Physics}, 8th edition (Wiley,
	  2004), chapter 13.
	\bibitem{Engelhardt2015}
	Larry Engelhardt, \enquote{Magnetic resonance: {Using} computer simulations and
	  visualizations to connect quantum theory with classical concepts,}
	  \href{http://dx.doi.org/10.1119/1.4930081}{American Journal of Physics}
	  \textbf{83}~(12), 1051--1056 (2015).
	\bibitem{Tung1985}
	Wu-Ki Tung, \emph{Group Theory in Physics} (World Scientific, Singapore, 1985).
	\bibitem{Friedrich2003}
	Bretislav Friedrich and Dudley Herschbach, \enquote{{Stern} and {Gerlach}:
	  {How} a bad cigar helped reorient atomic physics,}
	  \href{http://dx.doi.org/10.1063/1.1650229}{Physics Today} \textbf{56}~(12),
	  53--59 (2003).
	\bibitem{Wallis2013}
	Sean Wallis, \enquote{Binomial confidence intervals and contingency tests:
	  {Mathematical} fundamentals and the evaluation of alternative methods,}
	  \href{http://dx.doi.org/10.1080/09296174.2013.799918}{Journal of Quantitative
	  Linguistics} \textbf{20}~(3), 178--208 (2013).
	\bibitem{Note1}
	Statistically speaking, the simulation represents three independent Bernoulli
	  processes with probabilities $P_x(+1)$, $P_y(+1)$, and $P_z(+1)$, where the
	  values of the probabilities are related to each other by the normalization
	  constraint that the length of the Bloch vector equals one. This constraint
	  complicates the problem of estimating the state $\protect \mathbf {s}$ from
	  measurement data -- na{{\"i}}ve application of Eqs.~(\ref {eq:mean}) and
	  (\ref {eq:sig}) can yield nonphysical estimates of the state that violate the
	  normalization constraint $s_x^2+s_y^2+s_z^2 = 1$.
	\bibitem{Note2}
	The students were six sophomore students in a Modern Physics survey course and
	  eleven junior/senior students in an Introductory Quantum Mechanics course
	  taught from the spins-first perspective using Ref.~\onlinecite{McIntyre2012} as the primary textbook.
	\bibitem{Griffiths2004}
	David~J. Griffiths, \emph{Introduction to Quantum Mechanics}, 2nd edition
	  (Pearson Education, Harlow, 2004).
	\end{thebibliography}


\end{document}